 \documentclass{iopart}            
\usepackage{graphicx}
\usepackage{harvard}

\usepackage{iopams} 
\begin{document}

\title[On the angle between the first and the second Lyapunov vectors]
{On the angle between the first and the second Lyapunov vectors in spatio-temporal chaos}

\author{D Paz\'o, J M L\'opez and M A Rodr\'iguez}

\address{Instituto de F\'isica de Cantabria (IFCA), CSIC-Universidad de Cantabria, E-39005 Santander, Spain}
\begin{abstract}
In a dynamical system the first Lyapunov vector (LV) is associated
with the largest Lyapunov exponent and indicates 
---at some point on the attractor--- 
the direction of maximal growth in tangent space. The LV
corresponding to the second largest Lyapunov exponent generally points
at a different direction, but tangencies
between both vectors can in principle occur.
Here we find that the probability density function (PDF) of 
the angle $\psi$ spanned by the first and the second LVs
should be expected approximately symmetric around $\pi/4$ and
peaked at 0 and $\pi/2$. Moreover, for small angles we uncover
a scaling law for the PDF $Q$ of $\psi_l=\ln\psi$
with the system size $L$:  $Q(\psi_l)=L^{-1/2} f(\psi_l L^{-1/2})$.
We give a theoretical argument that justifies this scaling form
and also explains why it should be universal (irrespective of the system details)
for spatio-temporal chaos in one spatial dimension.
\end{abstract}


\section{Introduction}

The Lyapunov exponents are fundamental quantifiers of chaos \cite{Ott}. The directions in phase space
associated with them are generally referred to as the Lyapunov vectors (LVs). It is clear that the
Lyapunov vectors should play an important role, at least conceptually, in predictability
questions in meteorology \cite{Kalnay} and related sciences, or in achieving a microscopic description
of many particle systems \cite{morriss05}.

Since 2007 \cite{wolfe_tellus07,szendro07,ginelli07}, there has been a growing awareness
in the scientific community that
the set of vectors obtained as a byproduct of the standard method 
to compute the Lyapunov exponents \cite{benettin80} 
is not the most suitable way of defining the Lyapunov
vectors due to a series of artefacts these vectors exhibit. 
The so-called characteristic (or covariant) Lyapunov vectors (CLVs) are known since long time ago
\cite{eckmann} to be the only intrinsic (metric-independent) basis of LVs,
but only since 2007 their computation has become more or less routinary.
The use of these vectors has been probably more 
abundant in the field of meteorology due to their implication
in predictability questions \cite{legras96,trevisan98,pazo_tellus,crystal,sixto11}.

CLVs define the so-called Oseledec splitting or decomposition of tangent space. 
The concept of dominated decomposition is used in the mathematical
literature and basically implies that the Oseledec
subspaces are dynamically isolated. This is supposed not to be generically the case
in extended systems with spatio-temporal chaos \cite{yang08,yang09}.
However there is a lack of theoretical tools allowing
to know in advance how the angles between subspaces should be distributed.
Questions concerning the angles between the CLVs and the subspaces
they span have been addressed numerically in the context of
hydrodynamical Lyapunov modes \cite{yang08,bosetti10,morriss}, inertial 
manifolds \cite{yang09}, and hyperbolicity \cite{kuptsov10}.
In sum there is a growing interest on the angles among CLVs
in spatially extended chaotic systems, which
reflects in the latest publications on this subject \cite{morriss12}.

The seminal work by \citeasnoun{pik98} demonstrated that 
in extensive chaos, the first LV exhibits universal scaling laws
in space and time
falling into universality class of the KPZ equation \cite{kpz}.
Some system-independent scaling laws have been much more recently detected
for LVs corresponding to LEs smaller than the largest one \cite{szendro07,pazo08}.
This justifies the expectation that the angle between the LVs should
obey as well some universal features at a scaling level.
Eventually, the final picture of the
relations between different CLVs should be consistent
with the extensive nature of spatio-temporal chaos \cite{ruelle82,cross93}.

In this work we demonstrate that the probability density function (PDF)
of the angle between the two leading CLVs has universal features.
For small angle values, we uncover a 
universal (i.e.~system-independent) scaling law with the system size.
Our theoretical arguments make use of (i) the formulas intrinsic to
the method by \citeasnoun{wolfe_tellus07},
and (ii) the belonging (under a suitable transformation)
of the first LV to the universality class of the KPZ equation.

\section{Lyapunov vectors: Definitions}

In a $N$-dimensional dynamical system infinitesimal perturbations $\mathbf{\delta u}$
evolve governed by linear equations (the so-called `linear tangent model'). This implies the existence of
a linear operator $\mathrm{\mathbf{M}}$ that transforms the perturbation at a given time $t_1$
into the perturbation at another time $t_0$:
\begin{equation}
 {\mathbf {\delta u}}(t_0)=\mathrm{\mathbf{M}}(t_0,t_1) {\mathbf{\delta u}}(t_1)
\end{equation}
with the obvious properties $\mathrm{\mathbf{M}}(t_0,t_0)=\mathbb{I}$ and $\mathrm{\mathbf{M}}(t_1,t_0)=
\mathrm{\mathbf{M}}^{-1}(t_0,t_1)$. 

\subsection{Backward Lyapunov vectors}

The multiplicative ergodic theorem \cite{oseledec} (see e.g.,~\cite{eckmann}) establishes
the existence of a limit operator 
\begin{equation}
{\mathbf \Phi}_{b}(t_0) = \lim_{t_1 \to -\infty}
[{\mathrm{\mathbf{M}}(t_0,t_1)} {\mathrm{\mathbf{M}}^*(t_0,t_1)}]^{\frac{1}{2(t_0-t_1)}}
\nonumber
\end{equation}
where the asterisk denotes the adjoint operator, such that 
the logarithms of the eigenvalues of $\Phi_b$ are the LEs $\{\lambda_n\}_{n=1,\ldots,N}$.
By convention we assume $\lambda_n\ge\lambda_{n+1}$.
Note that, in contrast to the LEs, the operator $\Phi_b$ depends on the position
in the attractor (parametrized by $t_0$).
Moreover the metric determining the adjoint of $\mathrm{\mathbf{M}}$ is 
relevant (although irrelevant concerning the LEs).
Thus the eigenvectors of $\Phi_b$ form an orthonormal basis $\{\mathbf{b}_n(t_0)\}_{n=1,\ldots,N}$,
within the particular metric adopted. 
This set of eigenvectors, so-called
{\em backward} LVs~\cite{legras96}, serve to define a set of nested subspaces.
The first LV $\mathbf{b}_1(t_0)$ generates the straight line $S_1(\mathbf{x}_0)$ corresponding to 
infinitesimal perturbations at $\mathbf{x}_0=\mathbf{x}(t_0)$ that shrink as $\sim\exp(\lambda_1 t)$
as $t\to-\infty$. $\mathbf{b}_1(t_0)$ and $\mathbf{b}_2(t_0)$
define the plane $S_2(\mathbf{x}_0)$, such that the modulus of infinitesimal perturbations
initially inside $S_2(\mathbf{x}_0)$, but outside $S_1(\mathbf{x}_0)$, obeys  $\sim\exp(\lambda_2 t)$
as $t\to-\infty$. Recursively, we define a set of nested subspaces,
$$
S_1(\mathbf{x}_0) \subset S_2(\mathbf{x}_0) \subset \cdots \subset S_N(\mathbf{x}_0) = \mathbb{R}^N
$$
such that if $\mathbf{\delta u}\in S_n\backslash S_{n+1}$ then
$\lim_{t\to-\infty} t^{-1} \ln\|\mathbf{\delta u}(t)\|=\lambda_n$.
(In the case degenerate LEs exist, trivial modifications in the above expressions
have to be performed.) 
The backward LVs coincide with the orthonormal vectors
obtained as a byproduct of the standard algorithm via Gram-Schmidt orthogonalizations 
to compute the LEs~\cite{ershov98}.

\subsection{Forward Lyapunov vectors}

The Oseledec theorem can be also formulated in the opposite time limit, defining an operator
\begin{equation}
{\mathbf \Phi}_{f}(t_0) = \lim_{t_2 \to \infty}
[{\mathrm{\mathbf{M}}^*(t_2,t_0)} {\mathrm{\mathbf{M}}(t_2,t_0)}]^{\frac{1}{2(t_2-t_0)}}
 \end{equation}
such that the LEs are the logarithms of the eigenvalues of ${\mathbf \Phi}_{f}$
and the eigenvectors form an orthogonal basis, called the forward LVs $\{\mathbf{f}_n(t_0)\}_{n=1,\ldots,N}$.
These vectors are the counterpart of the backward LVs, but now indicating
the directions that will grow in the future with exponents $\lambda_n$.
Like with the backward LVs, the Gram-Schmidt procedure
can be used to obtain forward LVs, but now going backwards in time 
and using the adjoint (e.g., the transposed) Jacobian matrix.
As noted by \citeasnoun{legras96},
the use of the transposed Jacobian makes the forward LVs to come up with
the standard ordering. This means that  to obtain the first $n$ forward LVs
we need to integrate only $n$ perturbations.

For numerical purposes, note that computing forward LVs requires to be able to
trace backwards a certain trajectory. This can be done in three different ways:
\begin{enumerate}
\item[(i)] Storing a complete trajectory in the computer (ideally in the RAM memory).
This is adequate if the system is not invertible, or in time-delayed systems \cite{pazo10}.

\item[(ii)] Storing periodically the state of the system along the forward integration.
This allows to integrate the system backwards, rectifying the trajectory
periodically to cancel out the departure from the attractor (now, a repellor)
with exponent $-\lambda_N$. 

\item[(iii)] Integrating backward with a ``bit reversible'' algorithm. This procedure has no cost
of memory but it works only with Hamiltonian systems \cite{mauricio10}.
\end{enumerate}

\subsection{Characteristic (or covariant) Lyapunov vectors}

The CLVs $\{{\bf g}_n\}_{n=1,\ldots,N}$
form the only truly intrinsic set of Lyapunov vectors, and
we will refer to them hereafter simply as the Lyapunov vectors.
CLVs are the Floquet eigenvectors in the case of a periodic orbit,
they are independent of the definition of the scalar product, and the associated expansion
rates are recovered in both, future and past, limits:
\begin{equation}
\lim_{|t| \to \infty} t^{-1} \ln ||{\mathbf{M}}(t,t_0) {\bf g}_n
(t_0) ||= \lambda_n .
\end{equation}
This property entails covariance with the (forward and backward) dynamics:
\begin{equation}
{\bf g}_n(t) \propto {\bf M}(t,t_0) {\bf g}_n(t_0) .
\label{covariance}
\end{equation}
We use the symbol of proportionality ``$\propto$'' instead of ``$=$'' because the norm
and orientation of the vector is arbitrary.

\section{Computation of CLVs: Wolfe and Samelson formulas}

In 2007 \citeasnoun{wolfe_tellus07} put forward a method
to compute the CLVs from backward and forward LVs,
solving a linear set of equations (see below).
A remarkable feature of Wolfe and Samelson algorithm is that it contains formulas that
should allow to achieve some theoretical progress in questions 
so far tackled only numerically.

The $n$-th (characteristic) LV can be expressed as a linear combination
of the first $n$ backward LVs:
 \begin{equation}
 {\mathbf g}_n(t)=\sum_{i=1}^n y^{(n)}_i(t) {\mathbf b}_i(t)
 \end{equation}
Wolfe and Samelson found that in addition to the first $n$ backward LVs,
the coefficients $y_i^{(n)}$ require the computation of only the first $n-1$ forward LVs. (This is
a great advantage if $n$ is much smaller than the dimension $N$ of phase space.)
Thus the $n$-dimensional vector of coefficients ${\mathbf y}^{(n)}$ is solution of a equation of the form
\begin{equation}
 {\mathrm D} {\mathbf y}^{(n)} = {\mathbf 0}, 
\label{py0}
\end{equation}
where the $n \times n$ matrix $\mathrm D$ is equal to $\mathrm{P^TP}$.
Hence, as noticed by Kuptsov and Parlitz \cite{kuptsov12}, it suffices to solve the equation
\begin{equation}
 {\mathrm P} {\mathbf y}^{(n)} = {\mathbf 0} . 
\label{py02}
\end{equation}
where ${\mathbf 0}$ is the $(n-1)$-dimensional null vector, and
${\mathrm P}$ is a $(n-1)\times n$ matrix with elements
\begin{equation}
 P_{ij}= \left< {\mathbf f}_i \cdot {\mathbf b}_j \right> .
\end{equation}
$\left< \cdot \right>$ denotes the scalar product, and the vectors are assumed to be normalized:
$\left< {\mathbf b}_i \cdot {\mathbf b}_j \right>=\delta_{ij}
=\left< {\mathbf f}_i \cdot {\mathbf f}_j \right>$.
Equation \eref{py02} consists of $n-1$ equations for $n$ unknowns. This under-determination
is not a problem because (assuming the LE is not degenerate) there exist an obvious indetermination
in the modulus and sign of the LV. We impose $\sum_{i=1}^n [y_i^{(n)}]^2=1$,
and hence only the orientation of the vector is not specified.

\section{The angle between the first and the second Lyapunov vectors}

In this work we restrict our study to the angle $\psi$ between
the first and the second LVs: 
\begin{equation}
\psi=\measuredangle({\mathbf g}_1,{\mathbf g}_2) . 
\end{equation}
As the signs of ${\mathbf g}_1$ and ${\mathbf g}_2$ are arbitrary,
we are free to choose them restricting $\psi$ to the interval $[0,\case{\pi}{2}]$.
After some algebra we can obtain from \eref{py0} (or \eref{py02})
a relation between $\psi$ and the angles between the two leading backward LVs and the
main forward LV,
$\alpha=\measuredangle({\mathbf b}_1,{\mathbf f}_1)$ and 
$\beta=\measuredangle({\mathbf b}_2,{\mathbf f}_1)$:
\begin{equation}
\tan\psi= \frac{\cos \alpha}{\cos \beta}
\label{psi}
\end{equation}

In a high-dimensional systems forward and backward LVs are expected to be very seldom parallel.
(For instance, the angle $\phi$ between two random vectors in $\mathbb{R}^N$ is distributed
as $P(\phi)\propto \sin^{N-2}\phi$.)
Hence, the high dimensionality of phase space suggests to
work with the displacements from orthogonality:
\begin{eqnarray}
\delta \alpha &=& \frac{\pi}{2} - \alpha \\
\delta \beta &=& \frac{\pi}{2} - \beta
\label{delta}
\end{eqnarray}
Equation (\ref{psi}) may be written in these new variables:
\begin{equation}
 \tan \psi= \frac{\sin (\delta\alpha)}{\sin (\delta\beta)}
\label{psi2}
\end{equation}
In high-dimensional spaces the constraints of $\mathbf{b}_2$ are so weak
that we can legitimately expect $\delta\alpha$ and $\delta\beta$ to be very
similarly distributed. 
Notice that as a consequence, since $\tan(\case{\pi}{2}-\psi)=1/\tan\psi$,
$\psi$ should be in good approximation
distributed symmetrically around $\case{\pi}{4}$
(particularly if the system is large).
Moreover, as $\delta\alpha$ and $\delta\beta$ are expected to be near zero,
their quotient should favour values of $\psi$ close to 0 or $\frac{\pi}{2}$.
In fact a probability density function (PDF) of $\psi$ has been recently measured 
in numerical simulations of a quasi-one-dimensional system of hard disks
by \citeasnoun{morriss12} (see figure 18),
finding the aforementioned properties:
approximately symmetric around  $\case{\pi}{4}$ and peaks at 0 and  $\case{\pi}{2}$.
These features are also observed in our simulations (see below) in one-dimensional
systems with extensive chaos. Note though that eq.~\eref{psi2} is valid in any dimension. 

A much finer analysis is needed to understand the statistics of $\psi$, particularly
close to the tangency of leading Oseledec subspaces $\psi\to 0$.
This is studied in detail in section \ref{sec:PDF}.

\section{Numerical models}
\label{sec:pnr}
In this section we introduce the two systems we have numerically investigated,
and present our first numerical results.
The first one is a coupled-map lattice (CML), and the second one is a minimal
stochastic model of the LVs. Both systems were previously studied
by \citeasnoun{szendro07} and \citeasnoun{pazo08}, respectively, 
and are good test-bed systems with generic properties of spatio-temporal chaotic system.

\subsection{Coupled-map lattice}
Our one-dimensional CML reads:
\begin{equation} 
\label{cml}
u_j(t+1)= \epsilon\left\{f[u_{j+1}(t)]+f[u_{j-1}(t)]\right\}+(1 - 2\epsilon )f[u_j(t)],
\end{equation}
where the index $j$ runs from 1 to $L$ (the system size),
with periodic boundary conditions: $u_0(t)=u_L(t)$ and $u_{L+1}(t)=u_1(t)$.
Like in \cite{szendro07}, the coupling parameter
is chosen to be $\epsilon=0.1$, and $f$ is the logistic map $f(y)=4y (1-y)$.
With these parameters the system is hyperchaotic with $\lambda_n >0$ for $n/L < 0.795$.

\begin{figure}
 \centerline{\includegraphics*[width=120mm]{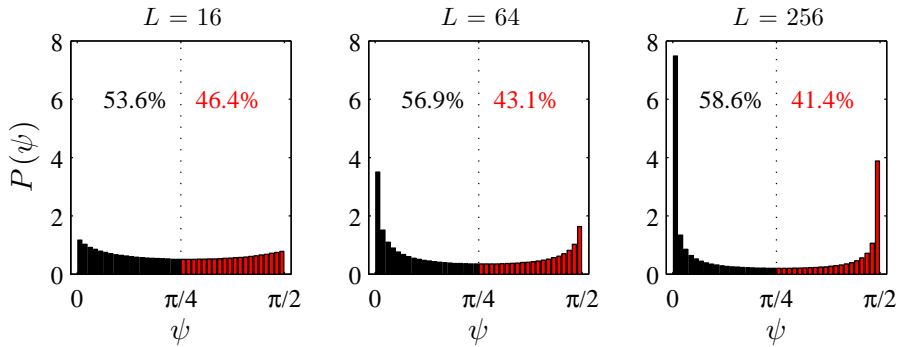}}
\caption{Distribution of $\psi$ for the CML \eref{cml} and three different system sizes.
The percentages refer to cumulative probabilities 
$\int_0^{\pi/4}P(\psi) \rmd\psi$ and $\int_{\pi/4}^{\pi/2}P(\psi) \rmd\psi$.}
\label{Fig:hist}
 \end{figure}

Our numerical simulations confirm that as anticipated in the previous section
the PDF of $\psi$ is roughly symmetric around $\case{\pi}{4}$,
see Fig.~\ref{Fig:hist}.
$P(\psi)$ is peaked at $0$ and $\case{\pi}{2}$.
However the distribution is not perfectly symmetric, and this unbalance
becomes more significant as the system size increases. Which is the asymptotic behaviour
of this unbalance as the system size increases will be a subject for future research.

Concerning the spatial organization of the CLVs, as
they are are known to be highly localized \cite{pik98,szendro07},
a probability of $\psi$ peaked at $0$, is consistent with an intermittent
coincidence of the localization sites of the first and second LVs. This is observed
in this CML \cite{szendro07} and other systems with spatio-temporal chaos
\cite{pazo08,mauricio10,sixto11}, and in time-delayed systems \cite{pazo10}.

\subsection{Minimal stochastic model}
\label{msm}
In two seminal works \citeasnoun{pik94} and \citeasnoun{pik98}
proposed the multiplicative stochastic linear equation
\begin{equation}
\partial_t w(x,t) = \zeta(x,t) w(x,t) + \partial_{xx} w(x,t) 
\label{noise}
\end{equation}
as a minimal model for the tangent space dynamic of
spatio-temporal chaos. $w$ represents the infinitesimal perturbation
and $\zeta$ is a stochastic forcing that mimics the chaotic forcing of the field.
$\zeta(x,t)$ is in general short-range correlated, and hence it
can be simply assumed to be zero-mean white noise with $\left< \zeta(x,t) \, \zeta(x',t')
\right>=2 \sigma \, \delta(x-x') \, \delta(t-t')$, as this assumption does not affect the
long-scale and long-time scaling properties of $w$.
Under a Hopf-Cole transformation, $h(x,t)=\ln|w(x,t)|$, equation \eref{noise}
becomes the KPZ equation \cite{kpz}
\begin{equation}
\partial_t h(x,t) = \zeta(x,t) + [\partial_{x} h(x,t)]^2 + \partial_{xx} h(x,t) ,
\label{kpz}
\end{equation}
which is a paradigmatic equation in the field of growing rough surfaces \cite{Barabasi}.
Under a Hopf-Cole transformation, the first LV falls into the universality class of the
KPZ equation. And thus, the large-scale spatial and temporal scaling properties 
of the LV are common to very different system types \cite{pik98,pazo10}, excluding Hamiltonian
lattices \cite{pik01} and disordered systems \cite{szendro08}.

In dynamical systems generic infinitesimal perturbations tend to align with the first LV,
whereas a measure zero set of perturbations may approach saddle-solutions of 
the linear tangent model, which are precisely the subleading LVs (i.e.,
corresponding to LEs smaller than the largest one).
In our previous work \cite{pazo08} we resorted to \eref{noise} as a minimal equation
for the sub-leading LVs. We assumed sub-leading LVs correspond to saddle solutions
of \eref{noise} for a given realization of the noise. 
In fact we observed that sub-leading LVs in systems with spatio-temporal chaos
and the saddle solutions of \eref{noise} display the same scaling and statistical properties. 
This similarity is realized after 
after taking the Hopf-Cole transformation of the LVs \cite{pazo08}. 

\begin{figure}
 \centerline{\includegraphics *[width=120mm]{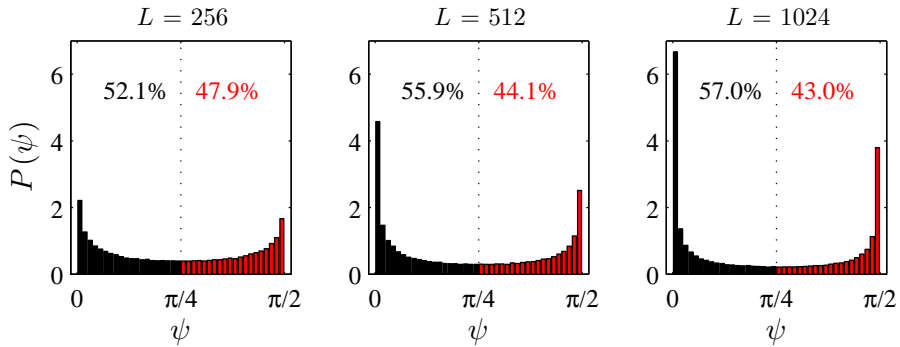}}
\caption{Distribution of $\psi$ for the minimal stochastic model \eref{noise}.}
\label{Fig:hist_ruido}
 \end{figure}

In our simulations of the minimal stochastic model
we have selected $\sigma=0.5$, like in \cite{pazo08}, for the variance of the noise.
Integration of several copies of \eref{noise} under
periodic orthonormalizations produces a set of vectors with the same spatio-temporal
structure than backward LVs \cite{pazo08} in a typical spatially extended chaotic system.
The equation for forward LVs is exactly the same as \eref{noise}
(because the operator in the right-hand side is self-adjoint).
This means that a meaningful $n$-th CLV can be computed from
the sets of $n$ backward and $n-1$ forward LVs obtained with the standard method \cite{benettin80}
integrating eq.~\eref{noise} with independent white noises $\zeta_b$ and $\zeta_f$, respectively.
Note that the obtained CLV at time $t_0$ is indeed a saddle solution of (\ref{noise})
for a particular realization of the noise: $\zeta=\zeta_b$ (for $t\le t_0$) and $\zeta=\zeta_f$ (for $t>t_0$).
The advantage of this procedure is that we can achieve good statistics
for the CLVs without the need for time-reversing the trajectory (i.e.~the noise).
The result of our numerical simulations is shown in figure \ref{Fig:hist_ruido}
and, as expected, exhibits the same qualitative features observe in
figure \ref{Fig:hist} for the CML.

\section{PDF of $\psi$ near the tangency ($\psi=0$)}
\label{sec:PDF}

To analyse the asymptotics of $P(\psi)$ for $\psi\to 0$,
it is useful to use the log-transformed variable
\begin{equation}
\psi_l \equiv \ln \psi ,
\end{equation}
denoted by the subscript $l$, likewise for $\delta\alpha_l$ and $\delta\beta_l$.
Let $Q$ be the PDF of $\psi_l$, with the trivial relation
\begin{equation}
Q(\psi_l)=P(\psi)\psi
\label{transfp}
\end{equation}

\begin{figure}
\centerline{\includegraphics *[width=120mm]{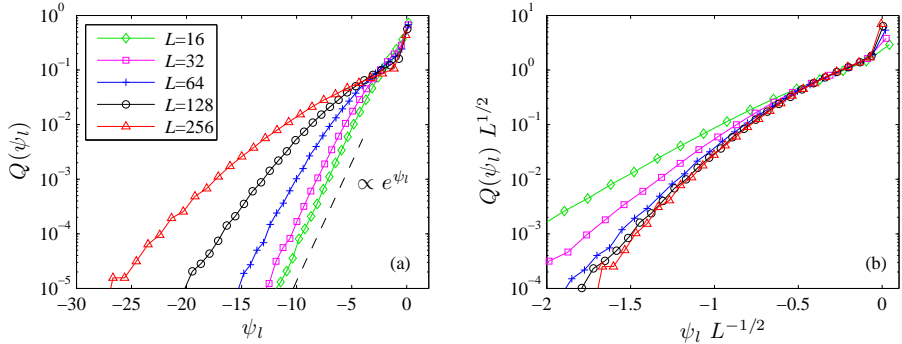}}
\caption{(a) PDF $Q$ of $\psi_l\equiv\ln\psi$ for
the CML, eq.~\eref{cml}. For small systems
the true asymptotic law $Q(\psi_l\to-\infty)\propto e^{\psi_l}$
can be detected.
(b) Data collapse via the scaling relation \eref{scaling_law}. The
region of data collapse progressively enlarges as the system size grows.}
\label{scaling_cml}
\end{figure}

\begin{figure}
\centerline{\includegraphics *[width=120mm]{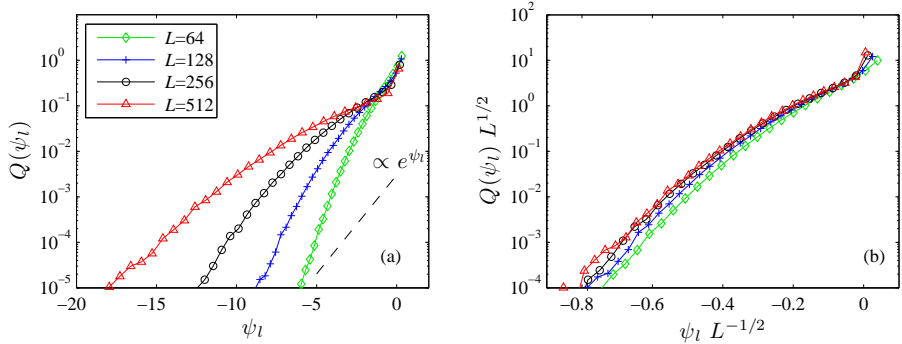}}
\caption{(a) PDF of $\psi_l$ for the minimal stochastic model \eref{noise}
and different system sizes. The decay as $\psi_l\to-\infty$ is faster
than $e^{\psi_l}$, and consistent with $P(\psi=0)=0$.
(b) Data collapse via the scaling relation \eref{scaling_law}. The
region of data collapse progressively enlarges as the system size grows.}
\label{scaling}
\end{figure}

The occurrence of tangencies ($\psi=0$) corresponds to a nonvanishing probability
at zero, i.e.~$\lim_{\psi\to0}P(\psi)= k >0$. Making a transformation
into variable $\psi_l$ this tranlates into an exponential dependence:
\begin{equation}
 \lim_{\psi_l\to-\infty}Q(\psi_l)= k e^{\psi_l}
\end{equation}
This asymptotic dependence in logscale is  observed in figure \ref{scaling_cml}(a) for the CML,
and we  conclude that $P(\psi=0)=k(L)>0$ with $k$ monotonically increasing with $L$.
Remarkably, despite the similarities observed between the CML and 
the minimal stochastic model, the latter behaves differently:
in figure \ref{scaling}(a) we may see that the PDF of $\psi_l$ decays faster than the exponential
for the minimal stochastic model. 
In fact we may proof that $P(\psi=0)=0$: in the minimal stochastic model
the first LV (either backward or forward because obey the same equation)
has the same sign in all the domain \cite{pik98,pazo08}, and in
consequence $\cos\alpha=\left< {\mathbf f}_1 \cdot {\mathbf b}_1\right>$
in eq.~\eref{psi} cannot vanish. 

The PDF of $\psi$ for the CML and the minimal stochastic model in
figures \ref{Fig:hist} and \ref{Fig:hist_ruido} look similar, but
figures \ref{scaling_cml}(a) and \ref{scaling}(a) evidence that the behaviour
of the PDF for $\psi\to0$ is very different in each model. 
We have found nonetheless that if $\psi$ is small but not extremely small,
there exist large enough values of $L$ such that the PDF $Q$ satisfies the scaling relation:
\begin{equation}
 Q(\psi_l)=L^{-1/2} f(\psi_l L^{-1/2})
\label{scaling_law}
\end{equation}
We may see in figures \ref{scaling_cml}(b) and \ref{scaling}(b) that 
there is a very good data collapse after scaling $\psi_l$ by $L^{-1/2}$.
Only when $\psi_l L^{-1/2}$ becomes smaller than a certain 
value $c(L)$ the peculiarities of each model show up.
Note that as $c(L)$ decreases with $L$ the departure from
the scaling law \eref{scaling_law} is not detectable for large systems.
Remarkably the departure from \eref{scaling_law} due to the finiteness of $L$
occurs upwards for the CML, and downwards for the minimal stochastic model,
reflecting their intrinsically different values of $P(\psi=0)$.

We emphasize that it is crucial to distinguish between the
limit $\psi\to0$ at large (but finite) $L$ and the limit $L\to\infty$ at
small (but nonzero) $\psi$. In the former case each model exhibits
its peculiarities and the stochastic model does not capture
the true behaviour for chaotic systems, 
which we expect to be generally like the CML. In fact, 
tangencies are believed to occur between ``physical'' modes \cite{yang09}.
However in the limit $L\to\infty$ the stochastic model captures the statistics of $\psi_l$
in the CML and presumably other chaotic systems.
Moreover we can justify the form of the scaling relation in eq.~\eref{scaling_law}
by virtue of some theoretical arguments that we develop in the next section.

\section{Theoretical analysis}

Close to tangency we can approximate $\tan\psi$ by $\psi$. In addition
we recall that $\delta\alpha$ and $\delta \beta$ can be expected to be close to
zero most of the time (as observed in the simulations, and not shown).
Hence in good approximation:
\begin{equation}
\psi= \frac{\delta\alpha}{\delta\beta}
\label{psi_aprox}
\end{equation}
In the log-transformed variables this relation becomes a subtraction
\begin{equation}
\psi_l = {\delta\alpha_l}-{\delta\beta_l}
\end{equation}

Let $G(\delta\alpha_l,\delta\beta_l)$ to denote the joint PDF
of $\delta\alpha_l$ and $\delta\beta_l$. An auxiliary variable
$\rho=\delta\alpha_l+\delta\beta_l$ allows to relate $Q$ and $G$ through the integral:
\begin{equation}
 Q(\psi_l)=\frac{1}{2}\int_{-\infty}^{\psi_l+2 \ln \frac{\pi}{2}}
G\left(\frac{\rho+\psi_l}{2}, \frac{\rho-\psi_l}{2} \right) \, \rmd\rho
\end{equation}
where we are assuming $\psi_l<0$.
$\delta\alpha$ and $\delta\beta$ are not completely
independent variables (e.g., in the seldom event
that one of them equals $\case{\pi}{2}$ the other one becomes 0).
However when both of them are close to zero ---which occurs most of the time---
we can expect them to be basically independent and the PDF factorizes:
$G(\delta\alpha_l,\delta\beta_l)\approx A(\delta\alpha_l) B(\delta\beta_l)$.
We get then
\begin{equation}
 Q(\psi_l)\approx\frac{1}{2}\int_{-\infty}^{\psi_l+2 \ln \frac{\pi}{2}}
A\left(\frac{\rho+\psi_l}{2}\right) B\left(\frac{\rho-\psi_l}{2} \right) \, \rmd\rho .
\label{integral}
\end{equation}
We are interested in the $\psi_l\to -\infty$ limit of this formula.

It is convenient to make a change of variable: $x=\case{\rho-\psi_l}{2}$, such that
\begin{equation}
 Q(\psi_l)\approx\int_{-\infty}^{\ln \frac{\pi}{2}}
A(x+\psi_l) B(x) \, \rmd x .
\label{integral2}
\end{equation}
This equation suggests that the asymptotics of the PDF of $\psi$ as $\psi\to0$ ($\psi_l\to-\infty$) is
highly influenced (if not determined) by $A$. So we focus our interest in the next section
on the distribution of $\delta\alpha$.

\subsection{PDF of $\delta\alpha$}

Recall $\alpha$ is the angle between the first backward LV and the first forward LV.
In a previous work \cite{pazo08} we found that
both vectors are well modeled by the multiplicative stochastic equation (\ref{noise}).
As we reasoned in Sec.~\ref{msm} the stochastic equation
can be used to get the statistics of characteristic LVs.
We have to integrate the fields $b_1(x,t)$ and $f_1(x,t)$, and the angle $\alpha$ between them
is obtained from a continuous version of the Euclidean scalar product (in practice the fields
are discretized so we compute the usual Euclidean scalar product):
\begin{equation}
 \cos\alpha=\frac{\left< b_1 \cdot f_1 \right>}{\left< b_1 \cdot b_1 \right>^{1/2}\left< f_1 \cdot f_1 \right>^{1/2}}=
\frac{\int_0^L b_1 f_1 \, \rmd x}{\left(\int_0^L b_1^2 \, \rmd x\right)^{1/2}  \, \left(\int_0^L f_1^2 \, \rmd x\right)^{1/2}} .
\end{equation}

A logarithmic transformation allows to define the associated surface for the first 
backward LV, $h_b(x,t)=\ln b_1(x,t)$,
and the first forward LV, $h_f(x,t)=\ln f_1(x,t)$.
Under this transformation $h_b$ and $h_f$ are governed by
the KPZ equation \eref{kpz}. We further decompose $h_b$ into the spatial average and the fluctuating
part $h_b(x)=\bar h_b + B_b(x)$ (and likewise for $h_f$). $\bar h_b$ is fixed by the norm of the vector and therefore
the result must be independent of the norm used. Some algebra yields the expression:
\begin{equation}
 \cos\alpha = \frac{\int_0^L \rme^{B_b(x)+B_f(x)} \, \rmd x}
{\left(\int_0^L \rme^{2B_b(x)} \, \rmd x\right)^{1/2}  \, \left(\int_0^L \rme^{2B_f(x)} \, \rmd x\right)^{1/2}}
\label{bbb}
\end{equation}
$B_b$ and $B_f$ are independent profiles {\em with zero mean}.
In particular, $B_b$ and $B_f$ are at long scales
indistinguishable from a Brownian path in one dimension, see \cite{pik98},
like solutions of the KPZ equation.
This kind of integrals in \eref{bbb} have been subject of some mathematical interest \cite{yor}
but unfortunately the theory is not mature yet as to provide results that 
one can readily use here, specially if different integrals are correlated.

Equation (\ref{integral2}) suggests small $\psi_l$ is controlled by the 
PDF asymptotic behaviour of $A(\delta\alpha_l\to-\infty)$, so we make the approximation
$\cos\alpha=\sin(\delta\alpha) \approx \delta\alpha$,
in the left hand side of
(\ref{bbb}). Next, taking the logarithms we obtain:
\begin{equation}\label{dal}
\delta\alpha_l \simeq \ln\left(\int_0^L \rme^{B_b(x)+B_f(x)} \, \rmd x\right)
-\frac{1}{2}\ln\left(\int_0^L \rme^{2B_b(x)} \, \rmd x\right)  -\frac{1}{2}\ln\left(\int_0^L \rme^{2B_f(x)} \, \rmd x\right)
\end{equation}

\subsection{Scaling with the system size}
From eq.~\eref{dal}, and recalling $B_b$ and $B_f$ are like two independent Brownian paths,
we can expect a scaling with the system size of the form $\delta\alpha_l \sim \sqrt{L}$.
This is expected to translate to $\psi_l$ in the form of the scaling relation in eq.~\eref{scaling_law}. 
We may also conjecture that in two spatial dimensions the scaling factor in \eref{scaling_law}
should be $L^{-\alpha_{2dKPZ}}$, $\alpha_{2dKPZ}\approx 0.387$ \cite{forrest90},
instead of $L^{-1/2}$.

We can also conjecture that in the minimal stochastic model
the PDF $P(\psi)$ vanishes not only at $\psi=0$ but in an interval
below a certain value $\psi^c$; i.e.~$P(0\le\psi<\psi^{c})=0$.
Note that in the most unfavourable situation,
if $B_b+B_f=0$, the numerator of \eref{bbb} equals $L$. And the denominator is maximal
if $B_b$ (and $B_f$) has a triangular shape. In this case the denominator grows
exponentially with $L$, and in turn $\delta\alpha_l$ is likely to have 
the infimum $\delta\alpha_l^{c}$ decreasing exponentially with $L$.
Hence $\delta\alpha_l^{c} \sim -L$, and looking
at eq.~\eref{integral2} we presume a similar dependence for the bound
for $\psi_l^{c} \sim -L$.

\section{Discussion and conclusions}

In the first part of this paper we have seen that Wolfe and Samelson formulas
allow to foresee the general form of $P(\psi)$, which should be general
in high-dimensional systems and irrespective
 of the dissipative or conservative character of the dynamics
and of the spatial dimensionality.
In the second part of the paper, we obtained a scaling law resorting
to a minimal stochastic model of the LVs for spatio-temporal chaos. 

The reason for the general validity of our scaling law \eref{scaling_law} is that
although specific system-dependent correlations between forward and backward LVs 
exist due to the deterministic nature of the dynamics,
these correlations are not expected to span much beyond the Lyapunov time $\lambda_1^{-1}$.
In contrast,  in typical systems with spatio-temporal chaos \cite{pik98}, see also \cite{pik01},
the backward (forward) LVs depend on the past (future) within large temporal range
of order  $L^z$ (with $z=\frac{3}{2}$ for the KPZ universality class in one spatial dimension).

Our scaling \eref{scaling_law} should be generally observed in spatially extended systems,
but the exact asymptotics of the PDF at extremely small $\psi$ is specific
for each system, which would be far beyond numerical capabilities already
for moderately large systems.
Our results should be taken into account in numerical experiments
with CLVs because if $L$ is large we may not detect what is
peculiar for each model, but just generic model-independent features.

So far the tools borrowed from surface roughning formalism have been probably
the most useful ones in providing theoretical results for the Lyapunov vectors
in spatio-temporal chaos.
This work underpins the might of this approach.

\ack
DP~acknowledges support by Ministerio de Econom\'ia y Competitividad (Spain)
through the Ram\'on y Cajal programme.
Financial support from
the Ministerio de Ciencia e Innovaci\'on (Spain) under projects
No.~FIS2009-12964-C05-05 and No.~CGL2010-21869 is acknowledged.

\end{document}